\documentclass[aps,preprint,twocolumn]{revtex4}
\usepackage{epsfig}
\usepackage{amsmath}
\usepackage{epsfig}
\usepackage{amssymb}
\begin{document}
\title{\bf Symmetric Fermi-type potential well} 
\author{Zafar Ahmed$^{1,*}$, Sachin Kumar$^2$, Tarit Goswami$^3$, and Sarthak Hajirnis$^4$}
\affiliation{$~^1$Nuclear Physics Division,  Bhabha Atomic Research Centre, Mumbai 400 085, India\\
$~^*$Homi Bhabha National Institute, Mumbai 400 094 , India	\\
	 $~^2$Theoretical Physics Section, Bhabha Atomic Research Centre, Mumbai 400 085, India
	\\
	$^3$Department of Computer Science and Engineering,Jalpaiguri Govt. Engineering College, Jalpaiguri, West Bengal 735102, India \\
	$~^4$Department of Physics, The Institute of Science, 
	 Mumbai 400032, India}
\email{1:zahmed@barc.gov.in, 2: sachinv@barc.gov.in,  3: taritgoswami456@gmail.com, 4: sarthakhajirnis500@gmail.com}   
\date{\today}
\begin{abstract}
We utilize the amenability of the Fermi-type potential profile in Schr{\"o}dinger equation to construct a symmetric one dimensional well as $V(x){=}{-}U_n/[1+\exp[(|x|{-}a)/b]], ~ U_n{=}V_n[1+\exp[-a/b]]$.  We define $\alpha=a/b, ~\beta_n {=}b\sqrt{2m U_n}/\hbar$, we find $\beta_n$ values for which critically the well has $n$-node half bound state at $E{=}0$. Consequently, this fixed well has $n$ number of bound states. Also we obtain a semi-classical expression ${\cal G}(\alpha,\beta)$  such that the Fermi well has either $[\cal G]$ or $[{\cal G}]+1$ number of bound states. Here $[.]$  indicates the integer  part. We also confirm the consistency of $\cal G$ with the number of s-wave neutron energy levels in a central ($x\in (0,\infty))$  Fermi potential well.
\end{abstract}
\maketitle
Historically, many phenomena of the microscopic world have been comprehended by hypothesizing that the smallest system is trapped in a potential $V(x)$.  Quantized energies of  bound states are obtained by solving the Schr{\"o}dinger equation 
\begin{equation}
\frac{d^2\psi(x)}{dx^2}+\frac{2m}{\hbar^2}[E-V(x)]\psi(x)=0,
\end{equation}
for  $V(x)$.  Textbooks in quantum mechanics demonstrate this by using  a square (rectangular) well of depth $V_0$ and width $2a$. It is found [1] that the effective radius  parameter  ${\cal G}'=2 \pi^{-1}\sqrt{\frac{2mV_0a^2}{\hbar^2}}$ of the square well determines the number of bound states as $[{\cal G}']+1$, where $[.]$ denotes integer part.

Here in this para, we attempt to give an almost exhaustive list of the one-dimensional exactly solvable  potential wells. The potential wells like harmonic, Morse and Eckart  are discussed in some textbooks [2]. The solvable Rosen-Morse potential well is available in Ref. [3]. Grendenshtein  [4] introduced a hyperbolic potential which is now called Scarf II. An interesting collection of exactly solvable potentials is available in [5]. In 1984, Ginocchio [6] proposed a versatile  potential well, one central potential [7]  was utilized as a one dimensional well [8].  The last two potentials belong to the Natanson [9] class. In all these cases, the eigenvalues $E_n$ are interesting explicit functions of $n$ and the potential parameters. The Gaussian potential well is not exactly and analytically solvable, employing interesting numerical solutions bound states of  this well have been discussed [10]. With the advent of interesting numerical packages, fast and accurate calculations involving even higher order functions are possible now. This extends the scope of studying some more  potential wells easily which are solvable even in terms of higher order functions such as Bessel and Hankel functions. In this regard two exponential potential models  have been proposed [11, 12] earlier. 

Inspired by the low energy scattering of neutron and proton, the concept of scattering-length and the formation of deuteron [13, 14], one can define a $n$-node Half Bound State (HBS) [15] $\psi_*$ at $E=0$, for a symmetric  well as $\psi_*(\pm \infty, W_n)= A$ (constant) such that
\vspace*{-0.3 cm}
\begin{multline}
\psi_*(x=0, W_n)=0 ~\mbox{or}~  \psi'_*(x=0, W_n)=0,\\ W_n=\sqrt{2 m V_{0n} a^2}/\hbar
\end{multline}
according to whether $n(\ge 1)$ is odd or even, respectively. Also  odd (even) $n$ defines an odd (even) parity HBS. It turns out that when in an square well $W_n=n\pi/2, n=1,2,3,..$ [15], $n$-node HBS exists at $E=0$ and the potential well has $n$ number of bound states. When $W_n$ is slightly increased from these critical values,  the well starts possessing one more bound state at an energy a little below $E=0$. A well has at most one HBS $\psi_*$ and its existence is critical.

Fl{\"u}gee's book [16] introduces a Fermi-type potential well as an interesting  solvable central potential for s-wave which is well known to represent the mean potential inside a nucleus [17] which is more realistic and phenomenological than an infinitely deep square well or the harmonic oscillator well. It is also called  Wood-Saxon potential or Saxon-Wood potential [18]. We propose to use this Fermi-type potential as a symmetric one dimensional potential well
\begin{equation}
V(x)=-V_0 \frac {(1+e^{-a/b})}{1+e^{(|x|-a)/b}}, \quad V_0,a,b>0.
\end{equation}
$V(0)=-V_0$, $V(\pm \infty)=0$ and in the limit when $b\rightarrow 0$, $V(x)$ is a rectangular well. In Fig. 1, three instances ($b=0.1,0.5,1$) of $V(x)$ are plotted for $V_0=5$ and $a=3$ to show that it  becomes rectangular well with rounded edge, it then spills out of the rectangle to become  bell-shaped, further it becomes  a well which is sharp around the origin and wide on the base.

In this article, we utilize the available exact analytic solutions [16] of the  Schr{\"o}dinger equation (1) for the Fermi well potential (3) to study its bound states and HBS. We find an expression for the effective parameter ${\cal G}(V_0,a,b)$ from semi-classical consideration for the bloated square well (3). In this well, we fix the values of $V_0$ and $a$ and vary $b$ to find that the Fermi well has equal or more number of bound states than that of the rectangular well. Students will find it interesting that any combination of $V_0, a, b$ leading to ${\cal G}$ will have the number of bound states as  $[{\cal G}]$,  or  $[{\cal G}]+1$. Further, we fix $\alpha=a/b$ and find values of $\beta_n$ so that the well (3) has $n$-node HBS at $E=0$ and hence $n$ number of bound states. We calculate the corresponding value of ${\cal G}$ and see that either $[{\cal G}]$ or $[{\cal G}]+1$ equals $n$. An analytic formula for semi-classical eigenvalues of the Fermi well will also found.

For a square well, it is often not realized that this exact quantum mechanical criterion [1] also comes from semi-classical quantization rule that at a discrete energy $E=E_n$ [1,2] 
\begin{multline}
\pi^{-1} \int_{x_1}^{x_2} dx~ \sqrt{\frac{2m}{\hbar^2}[E_n-V(x)]}{=}n+\frac{1}{2},
\end{multline}
where $x_1$ and $x_2$ are real classical turning points such that $V(x_1)=E_n=V(x_2)$. But in the case of square well these are $\pm a$.
Here, $n$ gives the quantum number of the discrete energy bound state. The eigenvalues $E_n$ obtained by (4) are only approximate. When a potential well vanishes asymptotically the 
value of $n$ corresponding to $E=0$ can be expected to give an excellent estimate of the number of bound states in the well. Therefore the integral
\begin{equation}
{\cal G}=\pi^{-1}\int_{-\infty}^{\infty} dx \sqrt{\frac{-2mV(x)}{\hbar^2}}
\end{equation}
gives us an effective parameter ${\cal G}$. We will see that for the Fermi well  (3) in all the cases discussed here,  there are $[{\cal G}]$ or $[{\cal G}]+1$ number of bound states. This dichotomy is due to the approximate nature of the semi-classical quantization (4).

For the square well ${\cal G}$ equals ${\cal G}'.$
The function ${\cal G}$ which is actually proportional to the area enclosed by $\sqrt{-V(x)}$
on the $x$-axis  for the symmetric Fermi well can be obtained as
\begin{multline}
\hspace{-.5 cm}{{\cal G}(V_0,a,b)}{=}{4 \pi^{-1} \beta \sinh^{-1}{e^{\frac{a}{2b}}}}, \beta{=}{\sqrt{\frac{2mU_0b^2}{\hbar^2}}} \hspace{-.25 cm}
\end{multline}
where $U_0=V_0[1+ e^{-a/b}].$
For  fixed values of $V_0$ and $a$,  ${\cal G}(b)$ can be seen to be an increasing function of $b$ justifying that the square well has the lesser number of bound states than that of the Fermi wells. For large values of $z$, we have $\sinh^{-1} e^z \sim z+\log 2$. In the limiting case when $b \rightarrow 0$, ${\cal G}(V_0,a,0)={\cal G'}$. In the sequel, we will use symbols $V_{0n}, U_{0n}$, and ${\beta}_n$ to mean that these parameters are fixed to have just $n$ number of bound states critically. If these are increased slightly, the well will have one more bound state slightly below  $E=0$.
\vspace*{-.7 cm}
\begin{figure}
	\centering
	\includegraphics[width=7cm,height=5 cm,scale=1.3]{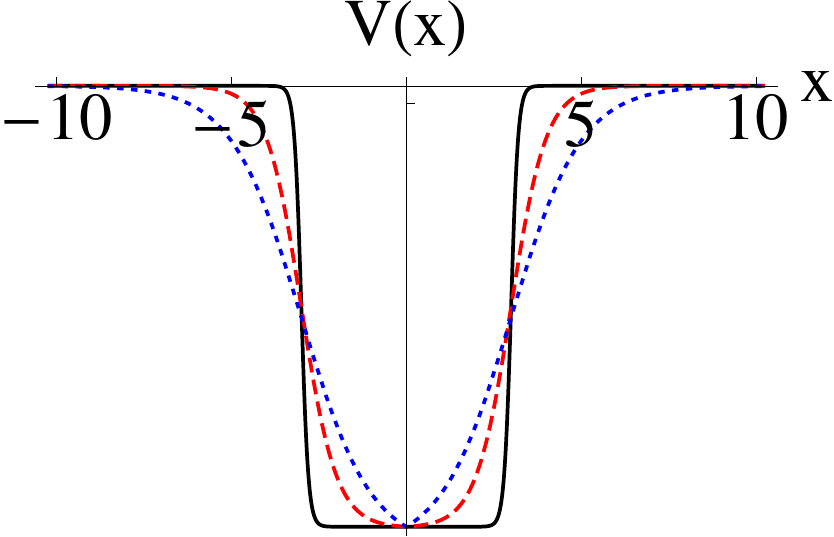}
	\caption{Three modifications of the Fermi well $V(x)$ (3). We fix $V_0=5, a=3$ and vary $b$ as 0.1 (black, solid), 0.5 (red, dashed), 1 (blue, short-dashed).}
\end{figure}

\begin{table}
	\centering
	\caption{Here $n$ denotes number of eigenvalues obtained using Eqs. (13) and (14), we show that irrespective of how a fixed value ${\cal G}$ is attained for various combinations of $V_0,a,b$
		the number of bound  states are the same: $[{\cal G}]$ or $[{\cal G}]+1$. \\ } 
	\label{my_label}
	\begin{ruledtabular}
		\begin{tabular}{ccccc}
			
			$\cal G$ & $a$ & $b$ & $V_0$ & $n$ \\
			3 & 1.5 & 0.9 & 48.6845 & 3 \\
			3 & 1.5 & 0.7590 & 60 & 3 \\
			3 & 1.0518  &  0.9 & 60 & 3 \\
			6.4 & 5 & 0.8 & 56.2945 & 6 \\
			6.4 & 5 & 0.6651 & 60 & 6 \\
			6.4 & 4.5090 & 0.7 & 70 & 6 \\
			8.7 & 6.8 & 0.7 & 64.4349 & 9 \\
			8.7 & 6 & 0.8646 & 75 & 9 \\
			8.7 & 6.0027 & 0.7 & 80 & 9 \\
			%	\label{tab:dis_size}   
		\end{tabular}
	\end{ruledtabular}
\end{table}

\section{Bound states}
\vspace*{-0.3 cm}
The Schr{\"o}dinger equation (1) for the Fermi potential (3) can be transformed by introducing [16]
 \begin{equation}
 \hspace*{-0.2 cm}y{=}\frac{1}{1+e^{\frac{(|x|-a)}{b}}}~\mbox{and}~ \psi(x)=y^{\nu}(1-y)^{\mu} \phi(y),
 \end{equation}
 to the Gauss hyper geometric equation [16,19]
\begin{multline}
 \hspace{ -.5cm} y(1{-}y)\phi''(x){+}[(2\nu{+}1){-}(2\nu{+}2\mu{+}2)y]\phi(x)\\{-}(\nu{+}\mu)(\nu{+}\mu{+}1) \phi(x)=0.
\end{multline}
\vspace*{-.8 cm}
\begin{figure}[h]
	\centering
	\includegraphics[width=3.5cm,height=4 cm]{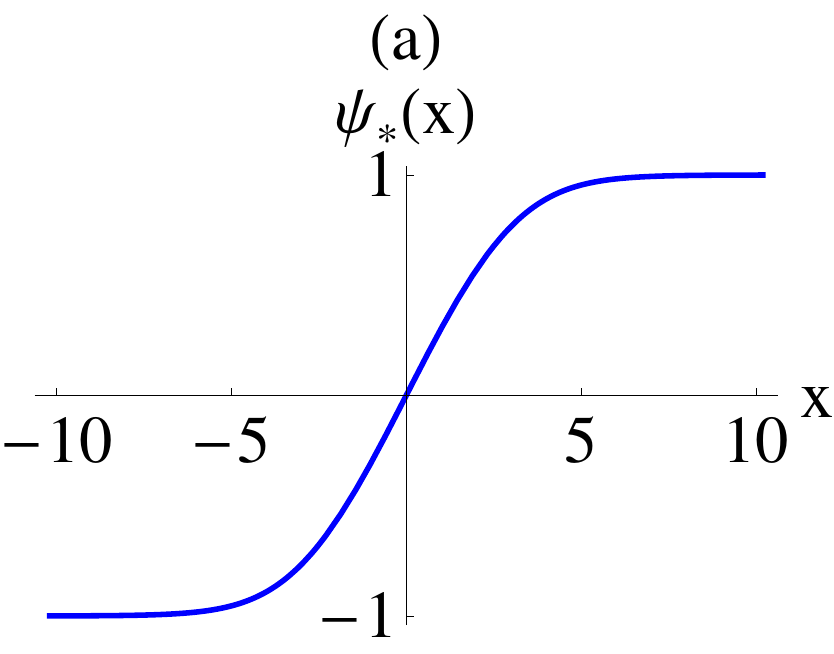}
	\hskip .5cm
	\includegraphics[width=3.5cm,height=4 cm]{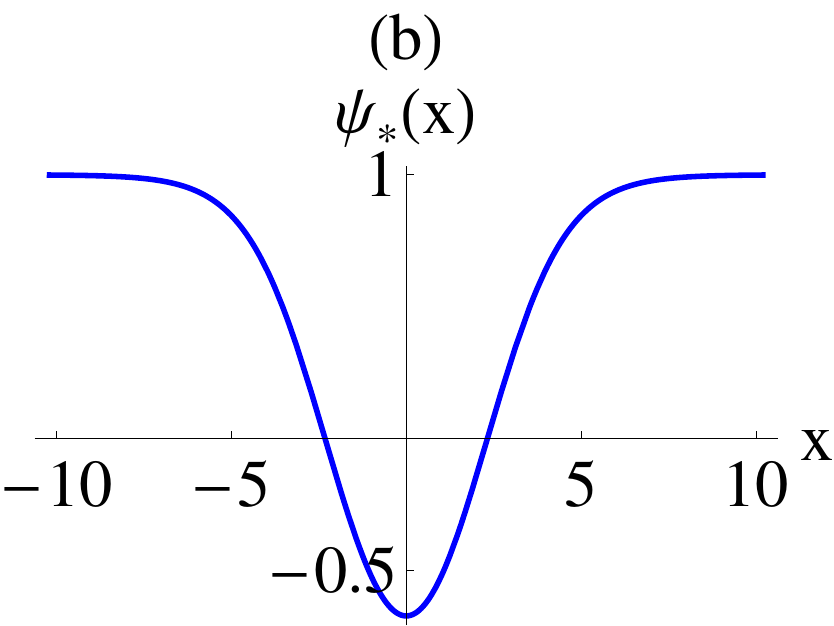}
	\hskip .5cm
	\includegraphics[width=3.5cm,height=4 cm]{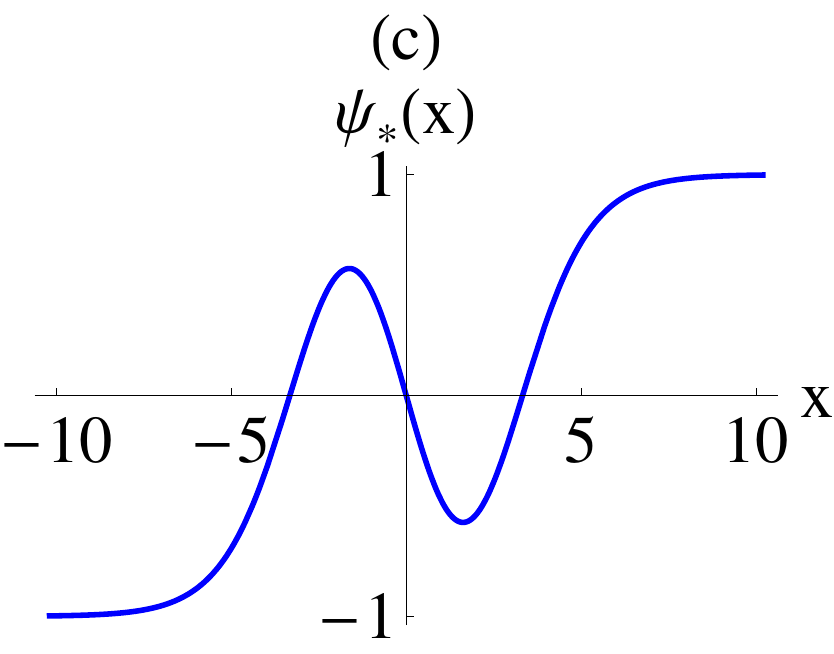}
	\hskip .5cm
	\includegraphics[width=3.5cm,height=4 cm]{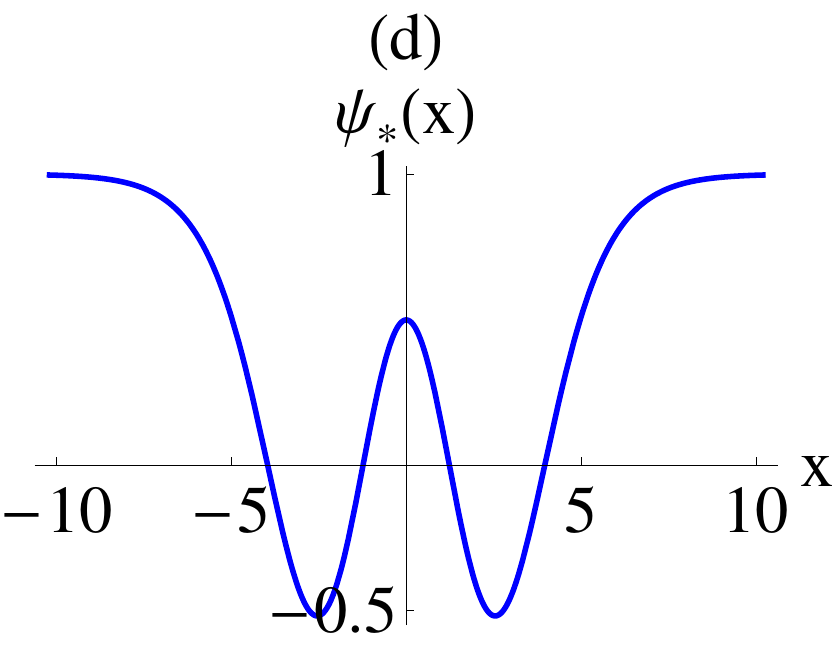}
	\caption {HBS $\psi_*(x)$ for symmetric Fermi well  potential (3) for $\alpha=4$ and $\beta_1=0.3697,\beta_2=0.6905,\beta_3=0.9947,\beta_4=1.2913$.}
\end{figure}

\begin{figure}[h]
	\centering
	\includegraphics[width=3.5cm,height=4 cm]{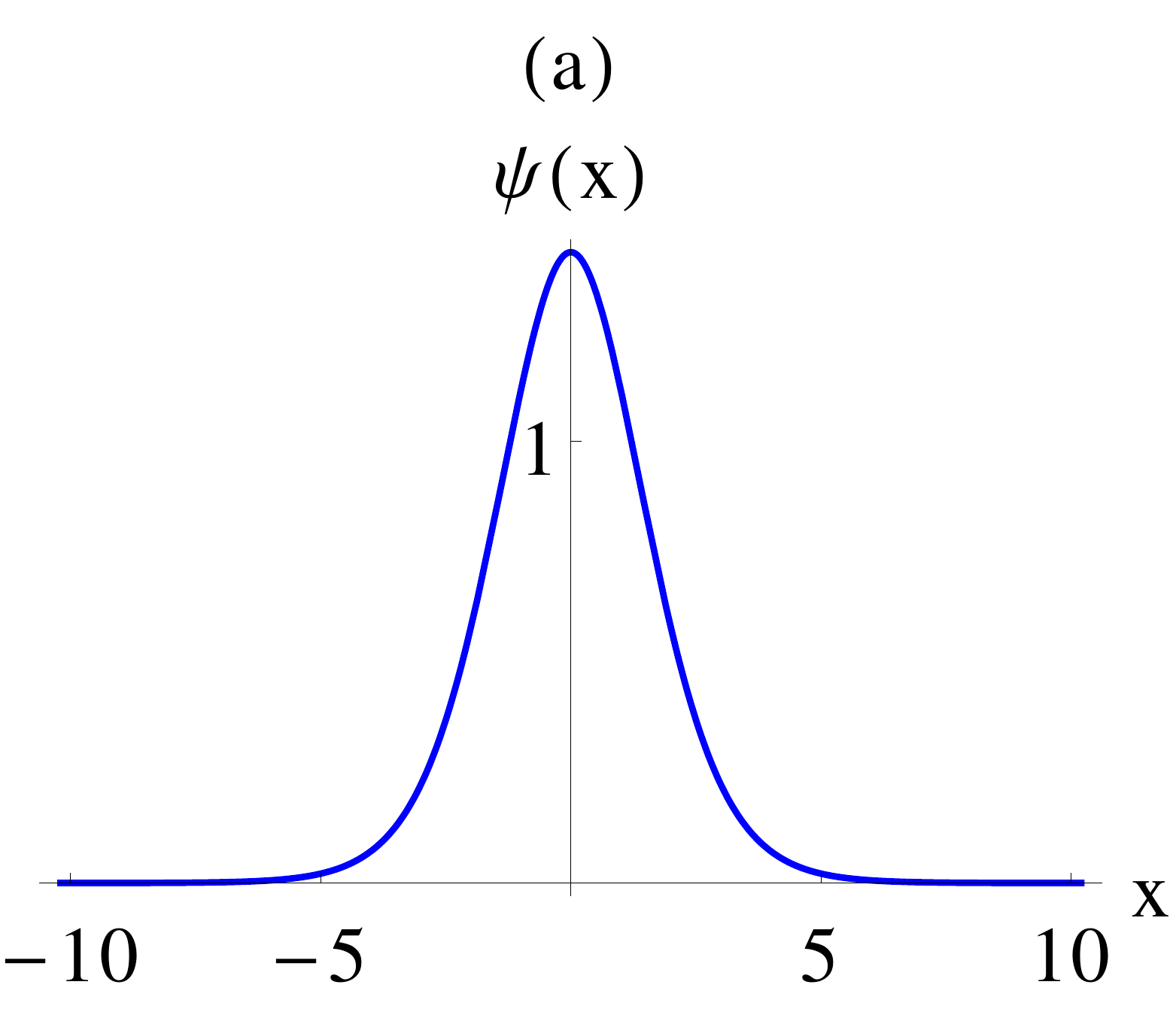}
	\hskip .5cm
	\includegraphics[width=3.5cm,height=4 cm]{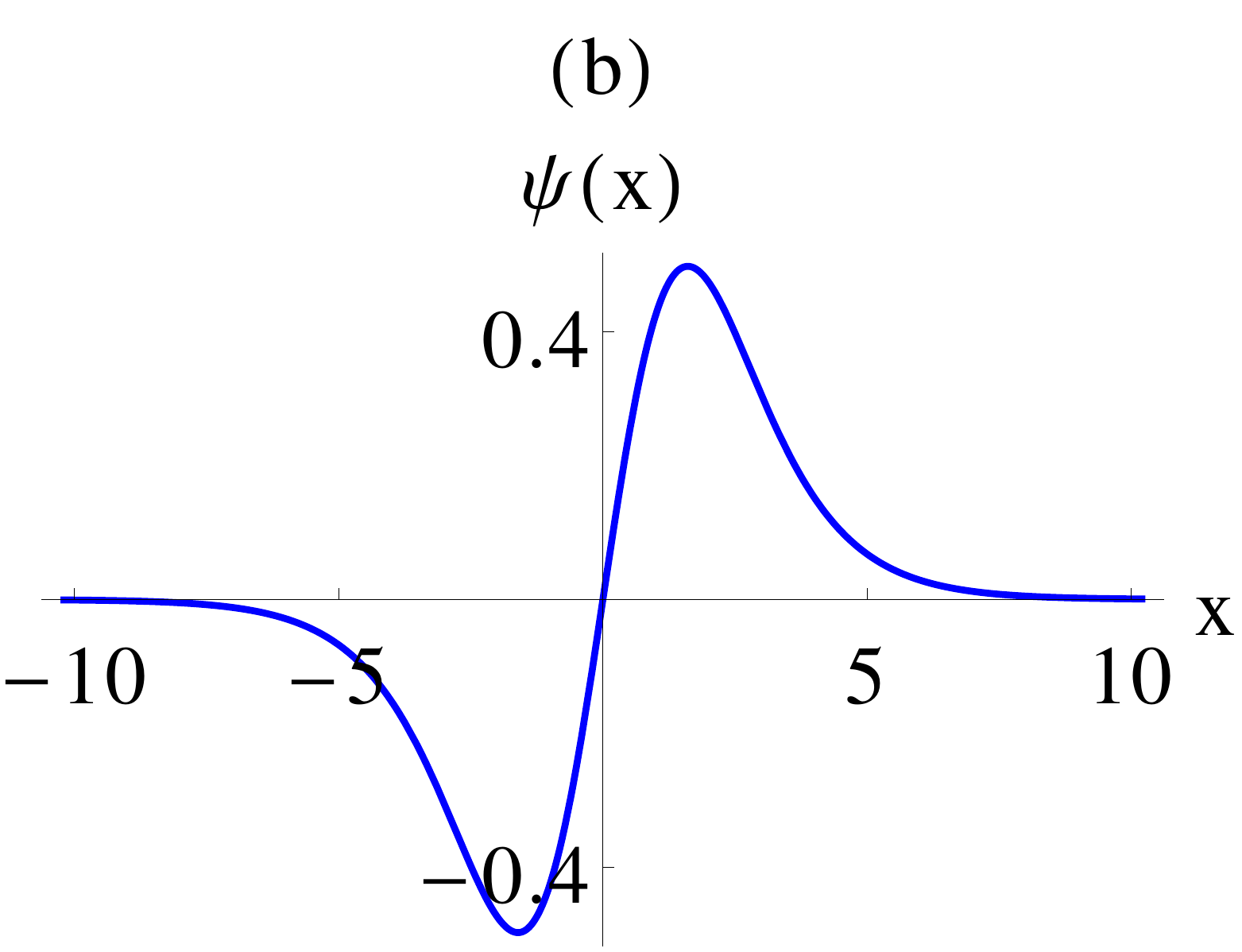}
	\hskip .5cm
	\includegraphics[width=3.5cm,height=4 cm]{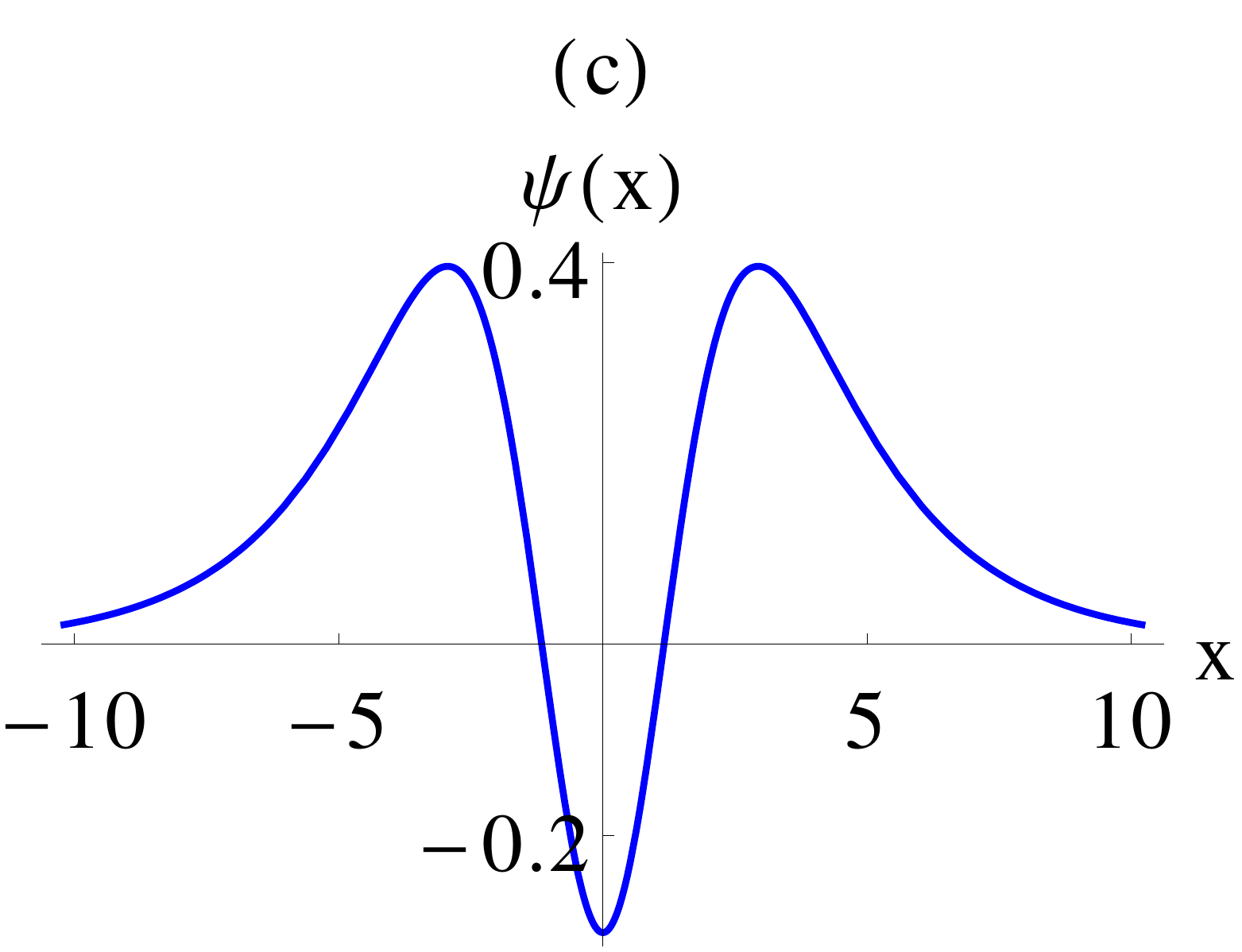}
	\hskip .5cm
	\includegraphics[width=3.5cm,height=4 cm]{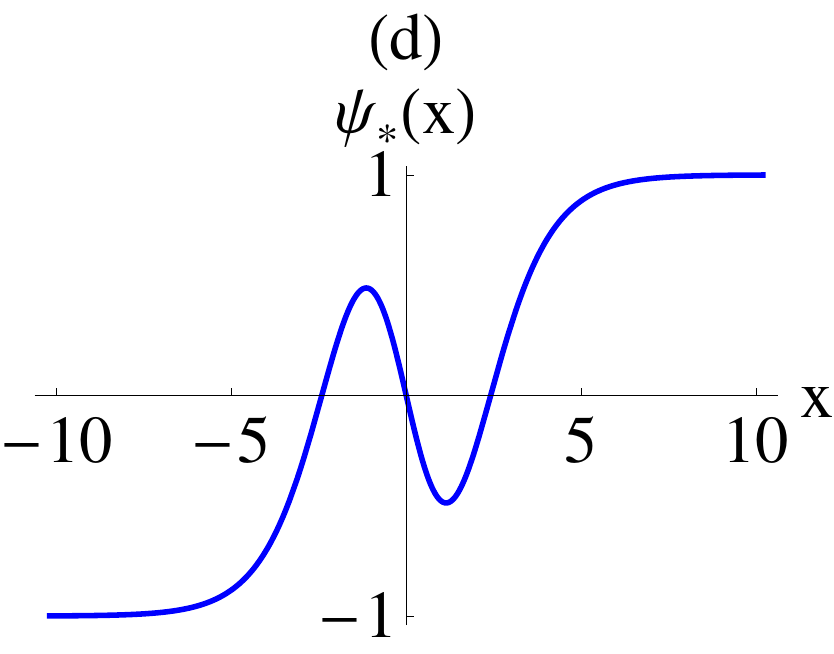}
	\caption {Three  bound states eigenfunctions $\psi_n(x)$ and the solitary HBS $\psi_*(x)$ of the well when $V_0=45.3642, a=2, b=1 (\alpha=2; \beta=1.5723)$. See the table II.}
	
\end{figure}
Here, we have $\nu=kb, \mu=ik'b$, with
\begin{equation}
k=\sqrt{\frac{-2mE}{\hbar^2}}~ \mbox{and} ~ k'=\sqrt{\frac{2m(E+U_0)}{\hbar^2}}.
\end{equation}
The second order linear differential equation (8) has two linearly independent solutions
\small 
\begin{multline}
\hspace*{-.4 cm}\phi_1(y){=} ~_2F_1[\nu+\mu,\nu+\mu+1;2\nu+1;y(x)], \\
\hspace*{-.45 cm} \phi_2(y){=} {\frac{1}{(1{-}y)^{2\nu}}}~_2F_1[\nu{-}\mu,\nu{-}\mu{+}1;1{-}2\nu;y(x)]. \hspace*{-.3 cm}  
\end{multline}
The infinite Gauss hyper geometric series for $~_2F_1[{\cal A},{\cal B};{\cal C};{\cal Z}]$ can be expressed as [16,19]
\begin{equation}
1+\frac{{\cal A} {\cal B}}{{\cal C} }\frac{{\cal Z}}{1!}+\frac{{\cal A}({\cal A}+1) {\cal B}({\cal B}+1)}{{\cal C}({\cal C}+1)} \frac{{\cal Z}^2}{2!}+...
\end{equation}
Notice that when $x \rightarrow \infty$, $y \sim e^{a/b} e^{-k|x|} \rightarrow 0$ and $~_2F_1 \rightarrow 1$. Therefore, the solution of (1) for (3) can be written as
\begin{eqnarray}
\psi(x,E){=}C y^{\nu}(1-y)^{\mu}~_2F_1[\nu{+}\mu,\nu{+}\mu{+}1; \nonumber  \\  2\nu{+}1;y],\hspace*{.2 cm}
\end{eqnarray}
which satisfies Dirichlet condition $\psi(\pm\infty)$ representing bound state correctly. On the other hand, the solution due to the other $~_2F_1(x)$ in Eq. (10) is unacceptable since it diverges as $e^{k|x|}$ when $x \sim \infty$. The potential (3) being symmetric, the solutions of (1) should be of definite parity. The even parity solutions are given as
\begin{multline}
\hspace{-0.5 cm}{\psi'(0,E_{2n})}{=}0,{~\psi_{2n}}{=}{\psi(|x|,E_{2n})},{~n{=}0,1,..}, \hspace{-.2 cm}
\end{multline}
and the odd parity solutions are characterized by
\small 
\begin{multline} 
\hspace*{-0.5 cm}{\psi(0,E_{2n+1})}{=}0,{~\psi_{2n+1}}{=}{\mbox{sgn}(x)}{\psi(|x|,E_{2n{+}1})}.\hspace{-.1 cm}
\end{multline}
For the Fermi potential the semi-classical eigenvalues can be obtained from Eq. (4) as
\begin{small}
	\begin{multline}
\hspace*{-0.5 cm}F(E){=}\frac{2\beta\sqrt{2}}{\pi}\Big[\sqrt{{\omega}{+}1}\tanh^{-1}\big(\frac{\sqrt{{ \omega}{+}\tanh(\alpha/2)}}{\sqrt{{\omega}{+}1}}\big) \\
	{-}\sqrt{{\omega}{-}1}\tanh^{-1}\Big(\frac{\sqrt{{\omega}{+}\tanh(\alpha/2)}}{\sqrt{{ \omega}{-}1}}\Big)\Big]{=}n{+}\frac{1}{2}, \\
	{\omega}(E){=}(1+\frac{2E}{U_0}),~ \alpha=\frac{a}{b},~\beta=b\sqrt{\frac{2mU_0}{\hbar^2}} .
	\end{multline}
\end{small}
\begin{table}
	\centering
	\caption{For different values of $\alpha$ and $\beta$ confirming the number of bound states to be $[{\cal G}]$ or one more than this. Here $\alpha$ and $\beta$ are dimensionless parameters defined in Eq. (15).\\ }
	\label{my_label}
	\begin{ruledtabular}
		\begin{tabular}{cccc|cccc}
			
			$\alpha$ & $n$ & $\beta_n$ & $\cal G$ & $\alpha$ & $n$ & $\beta_n$ & $\cal G$\\
			
			1 & 1 & 0.8774 & 1.4238 & 2 & 1 & 0.6226 & 1.3679\\
			& 2 & 1.4975 & 2.4302 && 2 & 1.1000 & 2.4166 \\
			& 3 & 2.1402 & 3.4731 && 3 & 1.5723 & 3.4541 \\
			& 4 & 2.7494 & 4.4617 && 4 & 2.0281 & 4.4555 \\
			& 5 & 3.3789 & 5.4833 && 5 & 2.4907 & 5.4716 \\
			& 6 & 3.9892 & 6.4735 && 6 & 2.9449 & 6.4694 \\
			& 7 & 4.6142 & 7.4878 && 7 & 3.4046 & 7.4794 \\
			& 8 & 5.2255 & 8.4798 && 8 & 3.8586 & 8.4767 \\
			\hline 
			3 & 1 & 0.4683 & 1.3150 & 4 & 1 & 0.3697 & 1.2700\\
			& 2 & 0.8534 & 2.3963 && 2 & 0.6905 & 2.3717 \\
			& 3 & 1.2234 & 3.4353 && 3 & 0.9947 & 3.4166\\
			& 4 & 1.5835 & 4.4465 && 4 & 1.2913 & 4.4354 \\
			& 5 & 1.9446 & 5.4604 && 5 & 1.5866 & 5.4496\\
			& 6 & 2.3018 & 6.4635 && 6 & 1.8796 & 6.4563\\
			& 7 & 2.6607 & 7.4713  && 7 & 2.1729 & 7.4636\\
			& 8 & 3.0172 & 8.4722 && 8 & 2.4650 & 8.4669
			\label{tab:dis_size}   
		\end{tabular}
	\end{ruledtabular}
\end{table}

\begin{table}
	\centering
	\caption{Testing the number of s-wave neutron levels in three nuclei in terms of $[{\cal G}/2].$ \\}
	\label{my_label}
	\begin{ruledtabular}
		\begin{tabular}{ccccccc}
			
			Element	& Mass number &$\cal G$ &  $n$ &$ n_{exp}$ \\
			O & 16 & 4.13 & 2 & 2 \\
			Sn & 132 &7.42 & 3 & 3 \\
			Pb & 208 & 8.49 & 4 & 4 \\
			%	\label{tab:dis_size}   
		\end{tabular}
	\end{ruledtabular}
\end{table}

\section{Half Bound States (HBS)}
\vspace{-0.3 cm}
For HBS, we set $E=0$ or $\nu=0$, $\mu=i\beta$ in Eq. (12) to get
\begin{align}
\psi_*(x,\beta)&=C (1-y)^{i\beta} ~_2F_1[i\beta, i\beta+1, 1,y] \nonumber \\  &=C~_2F_1[i\beta,-i\beta;1;y/(y-1)]
\end{align}
when $|x|\rightarrow \infty$, $y\rightarrow 0$, then $\psi_*(\infty) \rightarrow C$ which is nothing but the boundary condition on HBS (see above Eq. (2)). Next, the conditions that
\begin{equation}
\psi_*(0,\beta_n)=0 ~ \mbox{and} ~ \psi'_*(0,\beta_n) =0,
\end{equation}
are for odd and even node  solitary HBS of the well (3) at $E=0$, respectively.

In all the calculations here  we take $mc^2 \sim 940$ MeV,  $\hbar c \sim 197$ Mev fm, so $2m/\hbar^2=0.048 (\mbox{MeV fm}^2)^{-1}$. $V_0$ is in  MeV, $a$ and $b$ are in fm.
The Table I, is based on exact bound state eigenvalue calculations using Eqs. (13,14). Various combinations of $(V_0, a,b)$
giving rise to the same value of ${\cal G}$ (6) which allow the Fermi well (3) to possess $[{\cal G}]$ or $[{\cal G}]+1$ number of bound states. 

The Table II, is based on calculations for the solitary $n$-node HBS at $E=0$ of the Fermi wells using Eq. (17). We use the dimensionless effective parameters $\alpha$ and $\beta$. The critical $\beta_n$ values for four values of $\alpha$ have been calculated and the corresponding values of the effective parameter ${\cal G}$ are given. In Fig. 2, for $\alpha=4$, four $n$-node HBS for the critical values of $\beta_n (n=1,2,3,4)$ are presented. These four potentials are characterized with $\beta_1=0.30697, \beta_2= 0.6905, \beta_3=0.9947$, and $\beta_4=1.2913$ (see Table II),  The corresponding 
number of nodes are 1,2,3 and 4, and so are the corresponding number of bound states $[{\cal G}]$. An inaccurate value of $\beta$ may lead to the non-vanishing of $\psi_*(x)$ or its derivative at $x=0$, this would disturb the definite parity of the HBS.  If $\beta$ is slightly increased from $\beta_n$, the well will have $n+1$ bound states, the last one will be at an energy little below $E=0$.

In Fig.$~$3, three bound states and one HBS with three nodes are presented when $\alpha{=}2$ and $\beta{=}1.5723 (1.572333)$.$~$The plot of HBS is very sensitive to accurate value of $\beta$ which is root of Eq$.~(17)$.$~$Notice that 3-node HBS means three bound states in the potential well.$~$In Fig.$~$3, the eigenfunctions of three bound states ($E_0{=}-33.7554, E_1{=}-16.2221, E_2{=}-4.6764$) and the solitary  3-node HBS  at $E=0$  are presented. The corresponding semi-classical bound state eigenvalues obtained from Eq.$~$(15) are $-32.9723, -15.8589, -4.2151$, which are approximate but in a good agreement with the exact ones given above.

The half of the symmetric Fermi potential well (1) for $(x \in (0, \infty)$)   is also called the central Wood-Saxon potential [17,18] which represents, a nucleus. So the   odd levels of the full symmetric Fermi well (3) which are calculated by Eq. (13) are the same as $s$-wave levels of a nucleus. Therefore,  as per our criterion (5) the number of s-wave levels need to be  $[{\cal G}/2]$ or $[{\cal G}/2]+1$. For neutrons in a nucleus, we take  the typical parameters of Fermi well  as $V_0=50$ MeV, $a=1.3 (\mbox{A})^{1/3}$fm, $b=.65$ fm. Here A is the atomic mass number of a nucleus. For three nuclei: O, Sn  and Pb, we calculate ${\cal G}$ and the exact number of odd  eigenvalues using Eq. (13). We display this in the Table III, that  the half-well has either $[{\cal G}/2]$ or $[{\cal G}/2]+1$ number of eigenvalues, this in turn matches well the  experimentally observed number of $s$-wave levels in the respective nuclei.

Presently, we are not able to prove rigorously that number of bound states in a well which vanishes asymptotically in $(-\infty, \infty)$ are either $[{\cal G}]$ or $[{\cal G}]+1$ (5). However, in so many cases discussed here, it turns out to be true for the symmetric Fermi well in  three Tables presented here. Nevertheless, we did not find the number of levels to be $[{\cal G}]-1$, anytime. In this regard, an inventory for other one dimensional potential wells is most welcome.
%\vspace*{-0.7 cm}
\section*{\normalsize{CONCLUSION}}
\vspace*{-0.4 cm}
One dimensional symmetric  Fermi potential has been presented here as a
solvable potential well to study its bound states and the critical $n$-node half bound states. To the best of our knowledge, this as a one-dimensional potential well has been left out  in the literature so far. We hope that this potential model will be welcome as a new addition to the solvable one dimensional wells mentioned above. Moreover, our discussion of the number of bound states in this well semi-classically and in terms of $n$-node half bound state is instructive. Further, it will be interesting to study if the effective parameter ${\cal G}$ defined in Eq. (5) which comes quickly from the well known semi-classical quantization (6) determines the number of bound states  as well in other potential wells that vanish asymptotically on  one or both sides. 
\vspace*{-.2cm}
\section*{\normalsize{Acknowledgments}} We are thankful to the anonymous Referee for his/her several constructive suggestions.

\section*{\normalsize{REFERENCES}}
%\vspace*{-1.cm}

\end{document}